\documentclass[traditabstract]{aa} 
 
\usepackage{graphicx, times, txfonts, float, color, bm}
\usepackage{rotating}
\usepackage{longtable,lscape}
\usepackage[switch]{lineno}

\newcommand{\less}{\raisebox{-1.1mm}{$\stackrel{<}{\sim}$}} 
 
\newcommand{\msol}{\mbox{$M_{\odot}$}} 
 
\newcommand{\msolyr}{{$M_{\odot}$}\,yr$^{-1}$} 
\newcommand{\mdot}{$\dot{M}$}

\newcommand{\ks}{km s$^{-1}$}

\begin{document}

\title{
  Reflections on the photodissociation of CO in circumstellar envelopes
  %\thanks{
%Table~\ref{Tab-Res} is only available in electronic form at the CDS via 
%anonymous ftp to cdsarc.u-strasbg.fr (130.79.128.5) or via 
%http://cdsweb.u-strasbg.fr/cgi-bin/qcat?J/A+A/. 
%Figures~*** and *** are available in the on-line edition of A\&A. 
%}
%\fnmsep
%\thanks{
%Based on observations 
%} 
}  
 
\author{ 
 M.~A.~T.~Groenewegen\inst{1}
  \and
 M.~Saberi\inst{\ref{2},\ref{3}}
}

\institute{ 
Koninklijke Sterrenwacht van Belgi\"e, Ringlaan 3, B--1180 Brussels, Belgium \\ \email{martin.groenewegen@oma.be}
  \and
Rosseland Centre for Solar Physics, University of Oslo, P.O. Box 1029 Blindern, NO--0315 Oslo, Norway \label{2}
  \and
Institute of Theoretical Astrophysics, University of Oslo, P.O. Box 1029 Blindern, NO--0315 Oslo, Norway \label{3}
} 
 
\date{received: 2020, accepted: 2021} 
 
%\offprints{Martin Groenewegen} 
 
%\authorrunning{Groenewegen \& Bono} 
%\titlerunning{Photodissociation of CO in circumstellar envelopes} 
 
\abstract {
Carbon monoxide (CO) is the most abundant molecule after molecular hydrogen and is important for the chemistry in
  circumstellar envelopes around evolved stars.
  When modelling the strength and shape of molecular lines, the size of the CO envelope is an input parameter
    and influences the derived mass-loss rates.
In particular the low-J transition CO lines are sensitive to the CO photodissociation radius.
Recently, new CO photodissociation radii have been published using different formalisms that differ considerably.
One set of calculations is based on an escape-probability formalisms that uses numerical approximations derived
in the early-eighties.
The accuracy of these approximations is investigated and it is shown that they are less accurate
than claimed. Improved formalism are derived. Nevertheless, the changes in CO envelope size are small to moderate,
less than 2\% for models with $10^{-7}< \dot{M}< 10^{-4}$ \msolyr\ and at most 7\% for model with $\dot{M} = 10^{-8}$ \msolyr. 
}


\maketitle

\section{Introduction} 

  Essentially all low- and intermediate mass stars end their lives on the asymptotic giant branch (AGB), where they loose
  their enveloppe (of order $\sim$0.5 to 6.5\msol\ depending on inital mass) that is returned to the interstellar
  medium (ISM) in the form of dust and gas. Determining the mass-loss rate (MLR) is of key importance in understanding AGB evolution
  in detail. One of the main methods to do so is to use carbon monoxide (CO) as it is the most abundant molecule after molecular hydrogen and
  is very stable to photodissociation (see \citealt{HO18} for an overview). As CO also essentially binds all carbon when
  C/O$<$1 (oxgygen-rich stars, or O-stars) and all oxygen when C/O$>$1 (carbon-rich stars, or C-stars) the molecule is also important for
  the chemistry in the circumstellar envelopes (CSEs) around AGB stars, see e.g. \citet{Agundez20}.

  When performing detailed radiative transfer modelling of CO-lines (typically fitting multiple transitions) to determine the MLR
  (see e.g. \citealt{Sahai1990, Groenewegen1994, Schoier01, Decin06}) an CO abundance profile is required as input.
  The CO density is usually parameterised as
  \begin{equation}
    n_{\rm CO} \propto f_{\rm CO}  \exp\left( - \ln(2) \left( \frac{r}{R_{\frac{1}{2}}} \right) ^{\alpha} \right),
    \label{Eq-R}
\end{equation}
  where $f_{\rm CO}$ is the photospheric CO abundance, $r$ is the distance to the star, $R_{\frac{1}{2}}$ the distance where the
  CO abundance is half the photospheric value (and typicaly called the photodissociation radius), and $\alpha$ describes the shape
  of the profile.

Recently, \citet{Gr17} and \cite{Saberi19} presented results of photodissociation calculations of CO
  (i.e. values for $R_{\frac{1}{2}}$ and $\alpha$)  for
a large grid of MLRs, expansion velocities, $f_{\rm CO}$ values, and strenghts of the interstellar radiation field (ISRF).
Both papers were updates of \citet{MGH88} that was the classical reference for the CO photodissociation radii
in circumstellar envelopes for almost thirty years.
Both papers used different approaches and the CO photodissociation radii found by \citet{Saberi19} are 11-60\% smaller
than found by \citet{Gr17} \citep{Saberi19}.

\citet{Gr17} used the approach introduced in \citet{Li14, Li16}. It is based on the shielding functions of \cite{Visser09},
that depend on the column density of CO and molecular hydrogen (H$_2$)\footnote{For a choice of excitation temperature, Doppler width
  and $^{12}$CO/$^{13}$CO ratio.}, and a numerical integration scheme that takes into account the fact that at any
point in the wind UV photons can arrive from all directions (see Appendices~A in \citealt{Li14} and \citealt{Gr17}).
The shielding functions from \cite{Visser09} were derived for a static environment however.

\cite{Saberi19} have used the escape-probability formalism developed for an expanding envelope by \citet{MJ83} (herafter MJ83)
and later updated by \citet{MGH88}. They used the latest updated molecular data for CO from \cite{Visser09} and for H$_2$
to calculate the CO shielding functions. 

  Both models assume spherical symmetry, a smooth wind (that is no clumps), a constant velocity in the outflow, and
  simplified scattering and absorption properties of the dust in the 912 -- 1076~\AA\ region. 

In the escape-probability formalism the relevant integrations over angle are replaced by numerical approximations developed
in MJ83.
These approximations have been derived for a certain parameter space of line and continumm optical depths,
but neither \citet{MGH88} and \cite{Saberi19} checked whether these conditions are in fact met.

In this paper the validity of these approximations is investigated, and an improved formalism proposed.

The mathematical problem and numerical solution to this problem are presented in Sect.~2.
The calculations are presented in Sect.~3 and discussed in Sect.~4, while Sect.~5 concludes this paper.

\section{Equations and solutions}
Two equations are relevant in the description of the photodissociation process of CO.
The first is a term describing the continuum absorption of radiation,
\begin{equation}
  \gamma =  \frac{1}{2} \int_{0}^{\pi} d \phi \sin \phi \; \exp \left( -\tau_{\rm c} \; \phi/\sin \phi \right),
  \label{Eq-gam-ex}
\end{equation}
where $\phi$ is the angle with respect to the radius vector from the central star (MJ83) and its numerical approximation,
\begin{equation}
\gamma =  \exp \left( -1.644 \;  \tau_{\rm c}^{0.86} \right)
  \label{Eq-gam-ap}
\end{equation}
`with an accuracy of 1\% for $\tau_{\rm c} < 5$' (Eq.~5 in MJ83), where $\tau_{\rm c}$ is the continuum optical depth
measured radially outwards from a point at distance $r$ from the central star.

In MJ83 only the continuum absorption by dust was considered  $(\tau_{\rm d})$ but also the contribution in the
line wings of H$_2$ transitions $(\tau_{\rm H_2})$ should be included, 
$\tau_{\rm c} = \tau_{\rm d} + \tau_{\rm H_2}$ (see the discussion in \citealt{MGH88} and Eq.~A14 in \citealt{Saberi19}).

The second relevant equation describes the escape probability for a line photon,
\begin{equation}
\begin{aligned}
  \beta & = & \frac{1}{2} \int_{0}^{\pi} d \phi \sin \phi \; \exp \left( -\tau_{\rm c} \; \phi/\sin \phi \right) \phantom{xxxxxxx}  \\
        &   & \cdot  \left( 1 - \exp \left( - \tau_{\rm i} /\sin^2 \phi \right)\right) / \left( \tau_{\rm i} /\sin^2 \phi \right), 
\end{aligned}
  \label{Eq-bet-ex}
\end{equation}
which is approximated as
\begin{equation}
  \beta = \gamma \; [ \left(1 - \exp (-\alpha \; \tau_{\rm i}) \right) / (\alpha \; \tau_{\rm i})  - \Delta  ]
  \label{Eq-bet-ap}
\end{equation}
with $\alpha = 1.5$,
`where $\Delta$ is a small correction without which this expression for $\beta$ is accurate to within 10\%' (Eq.~8 in MJ83).
When
\begin{equation}
  \Delta = 0.20 \; \exp (-0.94 \; \tau_{\rm i}) \left(1 - \exp (- \tau_{\rm i}^{1.45} /1.5) \right) / \tau_{\rm i}, 
  \label{Eq-del-ap}
\end{equation}
Equation~\ref{Eq-bet-ap} `is accurate to $<0.5$\% if $\tau_{\rm d}$ = 0 and a few percent when
$\tau_{\rm d} \ne$ 0' (Eq.~9 in MJ83). Again, the line optical depth $\tau_{\rm i}$ is measured outwards from a point in the envelope
at distance $r$.

It should be pointed out that \citet{MGH88} and \cite{Saberi19} ignore the $\Delta$ term (see the Appendix
in \citealt{MGH88} and Eq.~A4 in \citealt{Saberi19}).
These two papers also used a different notation in the sense that the term between square brackets is denoted $\beta$ corresponding
to the CO self-shielding and they labeled $\gamma$ for CO-mutual shielding by H$_2$ and dust. 

It is remarked that Eqs.~\ref{Eq-gam-ex} and \ref{Eq-bet-ex} ignore the finite size of the central star.
Assuming that all radiation from behind the central star is blocked, and that the UV emission of the central star
itself may be ignored,
the integration over $\phi$ effectively runs from 0 to  $(\pi - \beta),$ where 
\begin{equation}
  \sin \beta = \frac{R_{\star}}{r}
  \label{Eq-star}
\end{equation}
indicates the angle subtended by the central star (or the inner radius of the envelope) from a point at distance $r$
(see Eq.~5 and Fig.~A1 in \citealt{Gr17}). This effect was taken into account in \cite{Gr17}, while the approximations in MJ83 assume
  that the central star is a point source.

The results in the next section have been obtained using routines written in Fortran77 from \cite{Press1992}:
a Romberg integration schema ({\it qromb}) to perform the numerical integrations\footnote{
The Fortran77 implementation is available from the authors for guidance.}, and
a nonlinear least-squares fitting routine (a Levenberg-Marquardt algorithm, {\it mrqmin})
to derive the coefficients in Eqs.~\ref{Eq-gam-ap}, \ref{Eq-bet-ap}, and \ref{Eq-del-ap}.
The precision of the numerical integration routine has been verified by comparing selected results
to those obtained using an on-line tool for such calculations\footnote{\url{https://www.integral-calculator.com/}}.

\section{Results}

\subsection{Calculation of $\gamma$}
\label{Sect-gam}

\begin{figure}

\centering
\includegraphics[width=0.91\hsize]{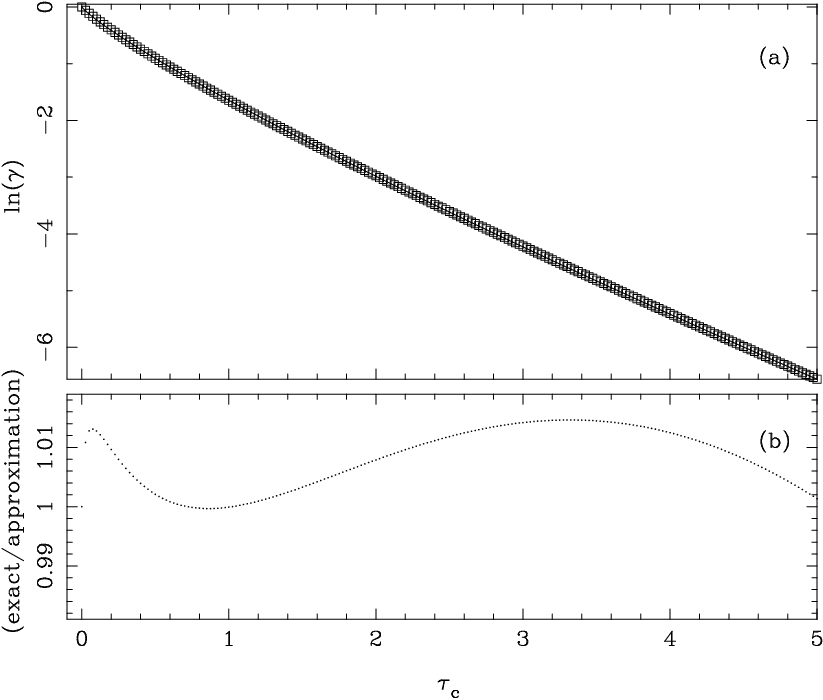}

\includegraphics[width=0.91\hsize]{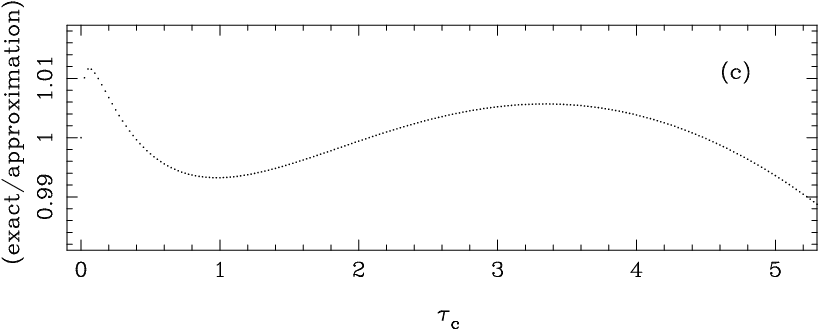}

\caption[]{ 
  Natural logarithm of $\gamma$ plotted against continuum optical depth (panel a), with the ratio of the exact solution to
  the approximation in MJ83, Eq.~\ref{Eq-gam-ap}, in (panel b).  Panel (c) show this ratio for the new best fit, Eq.~\ref{Eq-gam-better}.
} 

\label{Fig-gamma} 
\end{figure}

Figure~\ref{Fig-gamma} and Table~\ref{Tab-gamma} contain the results of the calculation of $\gamma$.
The upper two panels illustrate the results for $\tau_{\rm c} < 5$.
Panel (a) show the exact calculation of  $\gamma$ against $\tau_{\rm c}$. The approximation of Eq.~\ref{Eq-gam-ap} is also
plotted, but they are indistinguishable. Panel (b) shows the ratio of the exact calculation to that approximation.
The maximum deviation is $\approx 1.5\%$, larger than claimed in MJ83. One observes that in this range in optical depth
the approximation is systematically lower (by $\approx 0.5\%$) than the exact calculation.
For larger optical depths the deviations become increasingly larger, up to a factor of two
at $\tau_{\rm c} \approx 16$ (Table~\ref{Tab-gamma}).

As the numerical integration scheme is compared to a fitting formulae there is no observational error
as such to be used in a classical $\chi^2$ analysis.
Instead, an error will be assigned such that the final fitting formulae (Eqs.~\ref{Eq-gam-better2}, \ref{Eq-del-better2}) will
have a reduced $\chi^2$ of unity. This will allow to monitor the reduced $\chi^2$ in the several fitting steps
and will give representative error bars in the fitting coefficients.

Comparing in a chi-square sense Eq.~\ref{Eq-gam-ap} (with fixed coefficients $-1.644$ and $0.86$)
to the exact calculations over the range $0 < \tau_{\rm c} < 5$ in steps of 0.025 units
will result in a reduced $\chi^2$ of 7.3 if an `error' of 0.35\% (see below) is assigned to each data point.
Fitting for the coefficients and finding the solution with the smallest maximum deviation over the largest possible
range in $\tau_{\rm c}$ results in the best fit of:
 \begin{equation}
\gamma =  \exp \left( (-1.6375 \pm 0.0003) \;  \tau_{\rm c}^{(0.8618 \pm 0.0002)} \right),
  \label{Eq-gam-better}
\end{equation}
made in the range $\tau_{\rm c} < 5.3$. The maximum deviations are $\pm 1.2\%$ and the reduced $\chi^2$ becomes 2.0,
a significant reduction, indicating that the fit is much improved although the coefficients are similar.
  Panel (c) is a repeat of panel (b) for this approximation showing that the deviations are now more
symmetric around unity.

\begin{table}
\centering

\caption{Calculations for $\gamma$. }
%\footnotesize
  \begin{tabular}{cllllcclcccccc}
  \hline
$\tau_{\rm c}$ & $\gamma$  & exact/approx.  & exact/approx3.     \\
              & (exact)   &   Eq.~\ref{Eq-gam-ap}     &  Eq.~\ref{Eq-gam-better}          \\
  \hline
0.050 & 0.8938    &  1.013  & 1.012  \\
0.075 & 0.8486    &  1.013  & 1.012     \\
0.650 & 0.3215    &  1.000  & 0.995 \\
0.725 & 0.2874    &  1.000  & 0.994 \\
0.850 & 0.2393    &  1.000  & 0.994 \\
0.975 & 0.2001    &  1.000  & 0.993 \\
3.325 & 0.0010    &  1.015  & 1.006 \\
5.000 & 0.00142   &  1.001  & 0.994 \\ %no=201
5.300 & 0.00101   &  0.996  & 0.989 \\ %no=233
7.050 & 0.00014   &  0.950   & 0.946 \\
8.375 & 0.00003   &  0.900  & 0.899 \\
10.55 & 3.1 (-6)  &  0.800 & 0.802 \\
12.48 & 3.9 (-7)  &  0.700 & 0.705 \\
14.35 & 5.3 (-8)  &  0.600 & 0.609 \\
16.35 & 6.4 (-9)  &  0.500 & 0.509 \\

\hline
\end{tabular}
  \tablefoot{
Column~1 gives the continuum optical depth, Col.~2 the exact value of $\gamma$, and Cols~3 and 4 the ratio of the exact value to
    the approximate value according to Eqs.~\ref{Eq-gam-ap} and \ref{Eq-gam-better}, respectively. \\
    a (-b) stands for $a \cdot 10^{-b}$.
}

\label{Tab-gamma}
\end{table}

\subsection{Calculation of $\beta$}
\label{Sect-bet}

\begin{figure}

\centering
\includegraphics[width=0.75\hsize]{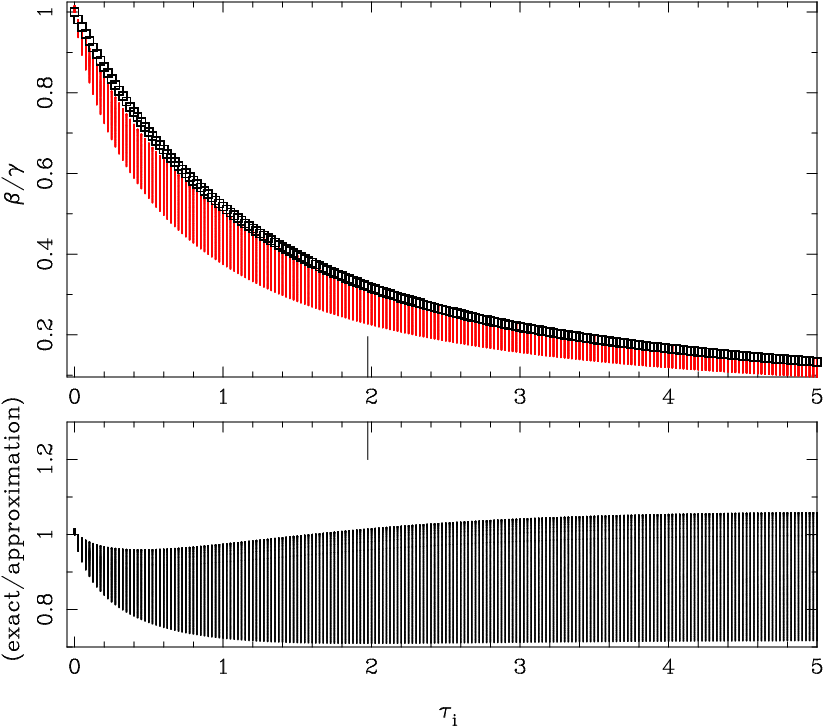}

\caption[]{ 
Ratio of $\beta$ over $\gamma$ plotted against line optical depth, for the approximation
in MJ83: Eq.~\ref{Eq-gam-ap} and Eq.~\ref{Eq-bet-ap} with $\alpha = 1.5$
and $\Delta = 0$ and calculated for $\tau_{\rm c} <5$.
The approximation is plotted with small squares in the upper panel.
The ratio of the exact solution to the assumed approximation is shown in the bottom panel.
For each discrete value of $\tau_{\rm i}$ (calculated at 0.025 intervals) the vertical lines
indicate the range in continuum optical depth, from 0 (largest value of $\beta$/$\gamma$) to the maximum value (lowest value).
The small vertical lines at the bottom of the upper and at the top of the bottom panel indicate
the optical depth where the difference between the exact calculation and the approximation is largest.
} 

\label{Fig-betagamma} 
\end{figure}

\begin{figure}

\centering

\includegraphics[width=0.75\hsize]{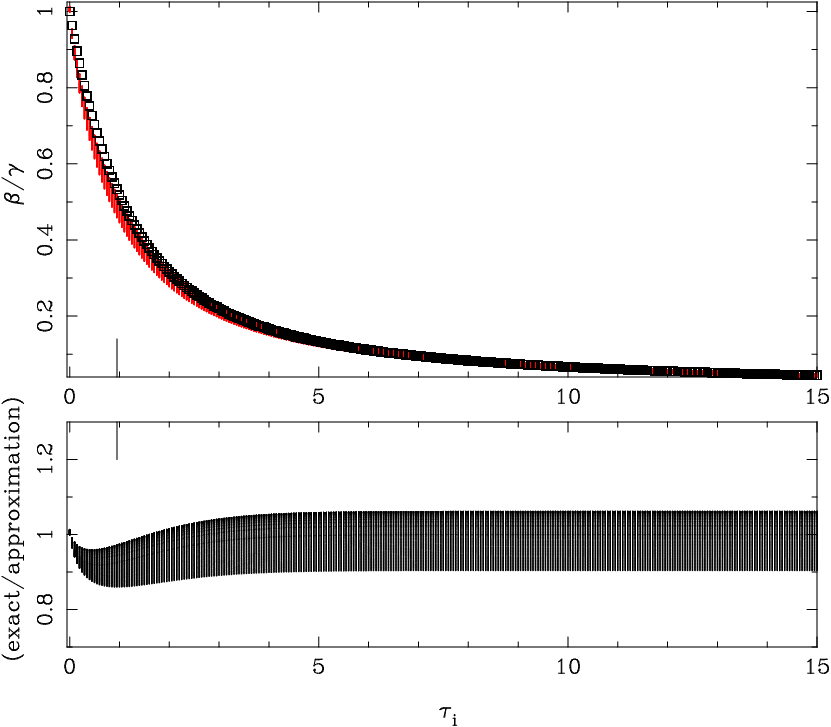}

\includegraphics[width=0.75\hsize]{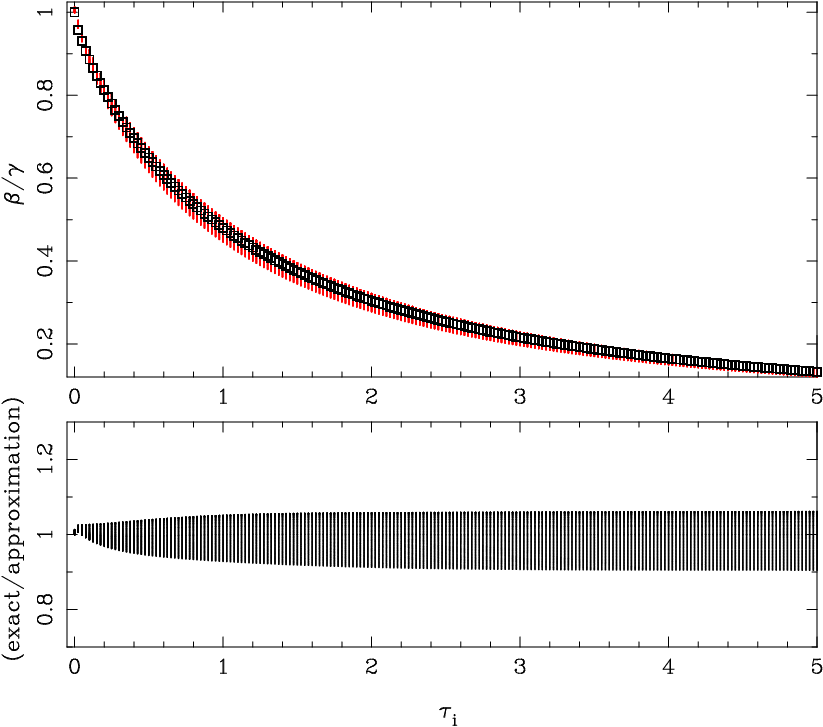}

\caption[]{ 
As Fig.~\ref{Fig-betagamma}, but with $\tau_{\rm c}$ limited to $<2.5$ (the top two panels). 
The calculations in the bottom two panels show the results with the $\Delta$ term included (Eq.~\ref{Eq-del-ap}).
that make the ratio of exact calculation to approximation more symmetric around unity.
} 

\label{Fig-betagamma1} 
\end{figure}

Figure~\ref{Fig-betagamma} compares the exact numerical results for the calculation of ($\beta/\gamma$) in the range
$\tau_{\rm i} < 5$ and $\tau_{\rm c} < 5$, to the approximation in Eq.~\ref{Eq-bet-ap} with $\Delta = 0$ and
$\gamma$ approximated by Eq.~\ref{Eq-gam-ap}.
The differences are much larger than the claimed 10\%, almost 25\% near $\tau_{\rm i} = 2.5$ (and for $\tau_{\rm c} = 5$).
Deviations of less than 10\% are only reached in a very limited range of $\tau_{\rm i}$ and $\tau_{\rm c}$.
Technically speaking, the claimed accuracies are on $\beta$, while $\beta/\gamma$ is considered here,
but as approximation Eq.~\ref{Eq-bet-ap}
is accurate to about 1.5\% this has little practical effect.
The deviations from unity appear to reach a constant level for large $\tau_{\rm i}$ and also depend on $\tau_{\rm c}$.
The top panels in Fig.~\ref{Fig-betagamma1} highlights this by showing the results for $\tau_{\rm i} < 15$ and
 the restricted range $\tau_{\rm c} < 2.5$.
At large optical depths the deviations are at most 10\%, but at $\tau_{\rm i} \sim 1.5$ it is 15\%.

The minimum in the ratio of exact-to-approximation curve near $\tau_{\rm i} \sim 1$
can be largely removed by including the term $\Delta$, as shown in the bottom panels Fig.~\ref{Fig-betagamma1}.
It shows the results for $\tau_{\rm i} < 5$, $\tau_{\rm c} < 2.5$, and using Eq.~\ref{Eq-del-ap} for $\Delta$.
The deviations from unity are now quite uniform with $\tau_{\rm i}$.

\begin{figure}

\centering
\includegraphics[width=0.75\hsize]{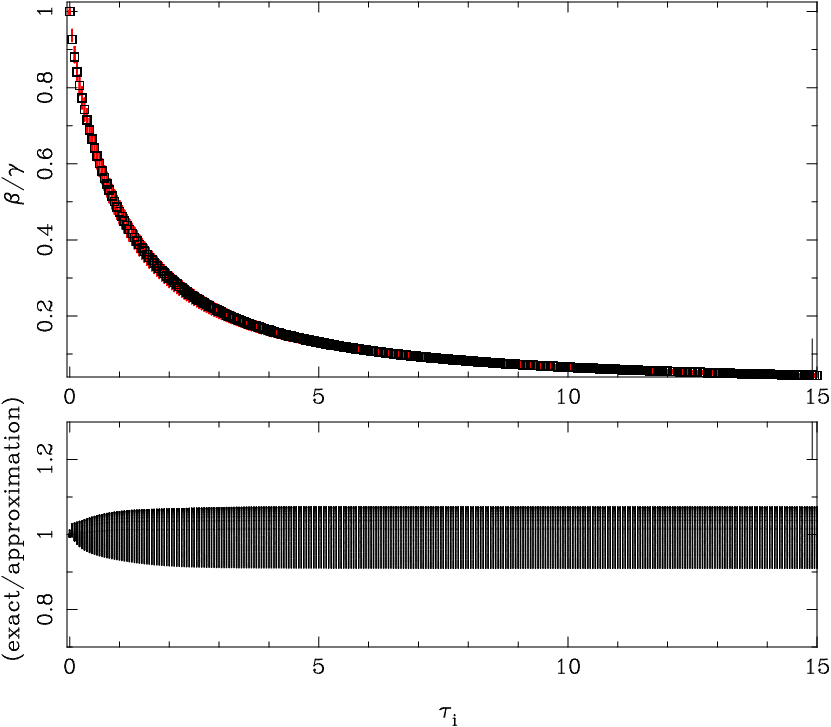}

\caption[]{ 
As Fig.~\ref{Fig-betagamma} for $\tau_{\rm c} <2.5$ using the new best fit: Eq.~\ref{Eq-gam-better},
Eq.~\ref{Eq-bet-ap} with $\alpha = 1.522$ and Eq.~\ref{Eq-del-better}.
} 

\label{Fig-betagamma2} 
\end{figure}

Taking the calculations with limits
$\tau_{\rm i} < 5$, $\tau_{\rm c} < 2.5$ and the approximations 
Eq.~\ref{Eq-gam-ap}, and
Eq.~\ref{Eq-bet-ap} with $\Delta = 0$ as reference (the default in \citet{MGH88} and \cite{Saberi19}),
the reduced $\chi^2$ is 45 when assigning an `error' of 1.0\% per datapoint (see below) in  $\beta/\gamma$.
Performing a fit in the range
$\tau_{\rm i} < 15$, $\tau_{\rm c} < 2.5$ and the better approximation
Eq.~\ref{Eq-gam-better} for $\gamma$, the best fit is
Eq.~\ref{Eq-bet-ap} with $\alpha = 1.5222 \pm 0.0001$ and
\begin{equation}
\begin{aligned}
  \Delta & = & (0.154 \pm 0.006) \; \exp (-(0.852 \pm 0.015) \; \tau_{\rm i}) \phantom{xxxxxx} \\
         &   &  \cdot   \left(1 - \exp (- (0.92 \pm 0.04) \; \tau_{\rm i}^{(1.441 \pm 0.006)} ) \right) / \tau_{\rm i}, 
\end{aligned}
  \label{Eq-del-better}
\end{equation}
with a reduced $\chi^2$ of 25. % (equivalent to an error of 5.0\% per datapoint).
This fit is shown in Fig.~\ref{Fig-betagamma2}.

\section{Average escape probabilities} 

In the previous section we showed that the approximations in MJ83 have limitations, but it is not clear how
this might affect a full calculation including all CO lines which cover a large range in optical depths.

To investigate this further, we considered the optical depths of the 855 CO lines for the standard model
in \citet{Saberi19}, with parameters
MLR \mdot = 10$^{-5}$ \msolyr, wind expansion velocity $V_{\rm exp}=$ 15~\ks,
$f_{\rm CO}$= 8 $\cdot 10^{-4}$, at a distance of $r= 10^{16}$~cm, which
implies a kinetic temperature of $\sim 70$~K (from Eq.~1 in \citealt{Saberi19}).
The CO line optical depth $\tau_{\rm i}$ ranges from 0.25 to $\sim 900$ with a median of 35.
The continuum optical depth $\tau_{\rm H_2}$ in the line wings of H$_2$ lines at the wavelengths of the CO lines
ranges from 0.12 to $>$10~000 with a median of 1.3.
The dust optical depth $\tau_{\rm d}$ is 4.14.

These optical depths scale like $\sim$ \mdot $f_{\rm CO}$/ $V_{\rm exp}$ /$r$.
\cite{Gr17} and \cite{Saberi19} considered MLRs from 10$^{-4}$-10$^{-8}$ \msolyr, expansion velocities from 3.75-30~\ks, and
CO abundances from 2-16 $\cdot 10^{-4}$.
The range in distances is taken from where the abundance (cf. Eq.~\ref{Eq-R}) changes from
(0.999 $\cdot$ $f_{\rm CO}$) to (2 $\cdot$ $R_{\frac{1}{2}}$).
Using the approximations in \citet{Gr17} for $R_{\frac{1}{2}}$ and $\alpha$ as a function of MLR, $f_{\rm CO}$, and $V_{\rm exp}$
it is determined that the optical depths range from about unity to $10^{-4}$ times those of the standard case
considering the entire parameter space.

Table~\ref{Tab-avebet} summarises the calculations for the average escape probability 
\begin{equation}
\overline{\beta} = \sum_{i=1}^{855} \; \beta_{i}
\end{equation}
over all CO lines for different scaling factors of the optical depth.
The escape probability is calculated: a) exactly, b) using the approximations in MGH and \citealt{Saberi19}
(Eq.~\ref{Eq-gam-ap} and Eq.~\ref{Eq-bet-ap} with $\alpha = 1.5$ and $\Delta = 0$), and c) the improved
approximations (Eq.~\ref{Eq-bet-ap} with $\alpha = 1.522$ and Eq.~\ref{Eq-del-better}), in Cols.~2, 3, and 4 respectively.

For the standard model the improved approximation is closer to the exact calculation, but still off by about 30\%.
In the standard model scaled downwards one observes the trend that the simple approximation systematically gives
too large average escape probabilities, while the improved approximation systematically gives lower values than the
exact calculation. The simple approximation actually gives results closer to the exact calculation, which is surprising
in view of the results presented in Sects.~\ref{Sect-gam} and \ref{Sect-bet}.

A closer inspection of the optical depths shows that in the standard case in 769 of the 855 lines
$\tau_{\rm i} > 15$ or $\tau_{\rm c} >5$ (and both conditions are violated in 365 lines), that is, values for the
optical depths outside the range of the fitting formula.

Considering the optical depth of all 855 lines for the 5 sets of models, and choosing escape probabilities
of $>$~0.001 as being the most relevant, we find that in 90\% of the cases the optical depths are
$\tau_{\rm i} < 7$ and $\tau_{\rm c} < 1$. This indicates that the line and continuum optical depths in realistic cases
  covers a more restricted range than considered in MJ83 and earlier in this paper.
Given this finding the fits to $\gamma$ and $\beta/\gamma$ were repeated using these limits, and the results are

\begin{equation}
\gamma =  \exp \left( (-1.6488 \pm 0.0013) \;  \tau_{\rm c}^{(0.8724 \pm 0.0015)} \right),
  \label{Eq-gam-better2}
\end{equation}
a value of $\alpha = 1.4376 \pm 0.0006$ in Eq.~\ref{Eq-bet-ap}, and
\begin{equation}
\begin{aligned}
  \Delta & = & (0.12 \pm 0.01) \; \exp (-( 0.75 \pm 0.04)\;  \tau_{\rm i}) \phantom{xxxxxx} \\
         &   &  \cdot   \left(1 - \exp (- (1.00 \pm 0.09)\;  \tau_{\rm i}^{(1.47 \pm 0.01)} ) \right) / \tau_{\rm i}. 
\end{aligned}
  \label{Eq-del-better2}
\end{equation}
   By earlier choosing the error per datapoint as 0.35\% and 1.0\%, respectively,
      the reduced $\chi^2$s are tuned to be unity in both fits. %  (equivalent to an error of 1.0\%).
With this new approximation the average escape probabilities (the last column in Table~\ref{Tab-avebet}) are very
close to the exact values, except in the standard case, where the functional forms of Eqs.~\ref{Eq-gam-ap} and \ref{Eq-bet-ap}
do not provide an adequate description.

\begin{table}
\setlength{\tabcolsep}{1.8mm}

\caption{Average escape probability calculated under different assumptions. }
%\footnotesize
  \begin{tabular}{lcccccccc}
  \hline
  Standard model & $\overline{\beta}$  & $\overline{\beta}$  & $\overline{\beta}$  & $\overline{\beta}$     \\
%                 &     (a)  &  (b)  &  (c)  &  (d)  \\ 
  \hline
$\cdot$ 1      & 4.22 (-5) &  5.73 (-5) & 5.56 (-5) & 5.04 (-5)  \\
$\cdot$ 0.1    & 8.18 (-2) &  8.35 (-2) & 7.84 (-2) & 8.16 (-2)  \\
$\cdot$ 0.01   & 0.492     &  0.507 & 0.478 & 0.491  \\
$\cdot$ 0.001  & 0.840     &  0.859 & 0.827 & 0.838   \\
$\cdot$ 0.0001 & 0.963     &  0.969 & 0.954 & 0.959  \\

\hline
\end{tabular}
  \tablefoot{
    Column~1 refers to the scaling factor of the dust, line and continuum optical depths relative to the standard model
    The average escape probabilities are calculated for:
    Col.~2, numerically exact;
    Col.~3, Eq.~\ref{Eq-gam-ap} and Eq.~\ref{Eq-bet-ap} with $\alpha = 1.5$ and $\Delta = 0$;
    Col.~4, Eq.~\ref{Eq-gam-better}  and Eq.~\ref{Eq-bet-ap} with $\alpha = 1.522$ and Eq.~\ref{Eq-del-better}; 
    Col.~5, Eq.~\ref{Eq-gam-better2} and Eq.~\ref{Eq-bet-ap} with $\alpha = 1.438$ and Eq.~\ref{Eq-del-better2}.\\
    a (-b) stands for $a \cdot 10^{-b}$.
    }
\label{Tab-avebet}
\end{table}

\begin{figure}
\centering
\includegraphics[width=0.95\hsize]{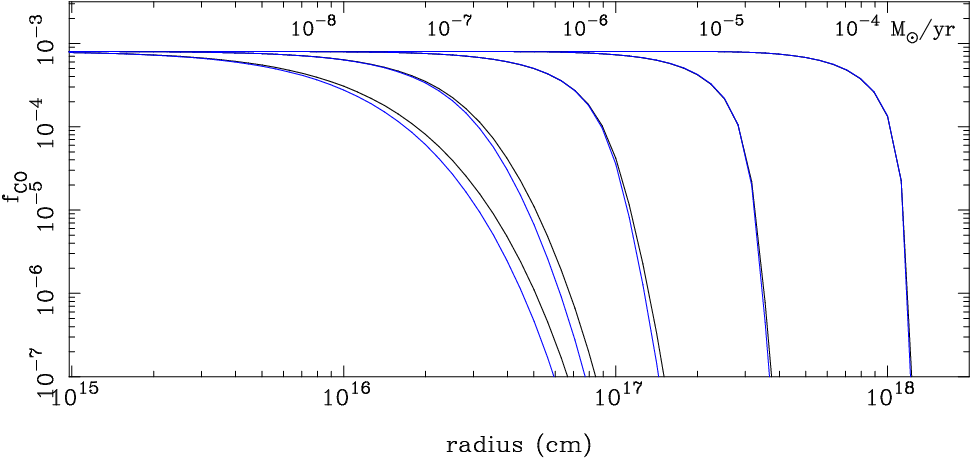}

\caption[]{ 
CO abundance profile for the 5 indicated MLRs.
Blue lines indicate the profiles using the approximations used in \cite{Saberi19},
black lines indicate the profiles using the improved approximations.
} 
\label{Fig-Res} 
\end{figure}

\section{Summary and discussion}

Improved numerical approximations are outlined to the formalism presented in MJ83 and that are at the basis of the CO
photodissociation calculations in \citet{MGH88} and \cite{Saberi19}.
The results in Table~\ref{Tab-avebet} show that the average escape probability in realistic cases deviate by 2-3\% from the exact
value for low to moderate
MLRs (and 35\% for large MLRs) in the approximation used by \citet{MGH88} and \cite{Saberi19}, and \less 0.4\% (19\%) in
the improved approximation.
For the quantity ($1-\overline{\beta}$) the differences are 0.002 - 19\%, respectively, 0.001 - 11\%, with the largest
difference for the smallest MLRs.

The improved approximations were implemented in the code used in \cite{Saberi19} and some calculations were performed using
$V_{\rm exp}=$ 15~\ks, $f_{\rm CO}$= 8 $\cdot 10^{-4}$, and MLRs \mdot = 10$^{-4}$, 10$^{-5}$, 10$^{-6}$, 10$^{-7}$, 10$^{-8}$ \msolyr.
Figure~\ref{Fig-Res} shows the CO abundance versus distance to the star for the 5 MLRs, from which
$R_{\frac{1}{2}}$ is determined. The improved formulae lead to a larger
photodissociation radius by 7.2\% for \mdot = 10$^{-8}$,  2.2\% for 10$^{-7}$ \msolyr, and \less 0.5\% difference for the other MLRs.
A larger photodissociation radius is expected as the escape probability is lower in the improved approximation compared
to the classical approximation.

As basic assumptions of spherical symmetry, constant velocity, and smooth outflow are identical in the \citet{Gr17}
  and \citet{Saberi19} models (see the introduction) the differences in the CO photodissociation radii (11-60\% smaller
  in \citet{Saberi19} than \citet{Gr17}) 
can not be attributed to them. These differences are due to the different underlying physical implementation, the CO shielding function
derived for a static environment (not representative of an AGB wind), and the escape-probability formalism in an expanding envelope,
respectively. It is shown here that the numerical approximations used in the latter formalism have a small effect on the outcome. 

The model by \citet{Saberi19} is currently the most accurate avialable and covers a large parameter space
(in MLR, $V_{\rm exp}$, $f_{\rm CO}$, ISRF strength). The results in the present paper show that the CO photodissociation radii are underestimated
by a few percent for the lowest MLRs in \citet{Saberi19}, but uncertainties in observational estimates of MLR and $f_{\rm CO}$ lead to larger
uncertainties.

Resolved observations of CO shells, like those carried out within the
DEATHSTAR\footnote{\url{www.astro.uu.se/deathstar}} project \citep{Ramstedt20} will allow for a detailed comparison of predicted and
observed CO photodissociation radii for a large sample in the near future.

\begin{acknowledgements}
MS acknowledges the SolarALMA project, which has received funding from the European Research Council (ERC) under the European Union’s Horizon
2020 research and innovation programme (Grant agreement No.  682462), and by the Research Council of Norway through its Centres of
Excellence scheme, project number 262622.
\end{acknowledgements}

\bibliographystyle{aa}
\bibliography{references}

\begin{thebibliography}{15}
\expandafter\ifx\csname natexlab\endcsname\relax\def\natexlab#1{#1}\fi

\bibitem[{{Ag{\'u}ndez} {et~al.}(2020){Ag{\'u}ndez}, {Mart{\'\i}nez}, {de
  Andres}, {Cernicharo}, \& {Mart{\'\i}n-Gago}}]{Agundez20}
{Ag{\'u}ndez}, M., {Mart{\'\i}nez}, J.~I., {de Andres}, P.~L., {Cernicharo},
  J., \& {Mart{\'\i}n-Gago}, J.~A. 2020, \aap, 637, A59

\bibitem[{{Decin} {et~al.}(2006){Decin}, {Hony}, {de Koter}, {Justtanont},
  {Tielens}, \& {Waters}}]{Decin06}
{Decin}, L., {Hony}, S., {de Koter}, A., {et~al.} 2006, \aap, 456, 549

\bibitem[{{Groenewegen}(1994)}]{Groenewegen1994}
{Groenewegen}, M.~A.~T. 1994, \aap, 290, 531

\bibitem[{{Groenewegen}(2017)}]{Gr17}
{Groenewegen}, M.~A.~T. 2017, \aap, 606, A67

\bibitem[{{H{\"o}fner} \& {Olofsson}(2018)}]{HO18}
{H{\"o}fner}, S. \& {Olofsson}, H. 2018, \aapr, 26, 1

\bibitem[{{Li} {et~al.}(2016){Li}, {Millar}, {Heays}, {Walsh}, {van Dishoeck},
  \& {Cherchneff}}]{Li16}
{Li}, X., {Millar}, T.~J., {Heays}, A.~N., {et~al.} 2016, \aap, 588, A4

\bibitem[{{Li} {et~al.}(2014){Li}, {Millar}, {Walsh}, {Heays}, \& {van
  Dishoeck}}]{Li14}
{Li}, X., {Millar}, T.~J., {Walsh}, C., {Heays}, A.~N., \& {van Dishoeck},
  E.~F. 2014, \aap, 568, A111

\bibitem[{{Mamon} {et~al.}(1988){Mamon}, {Glassgold}, \& {Huggins}}]{MGH88}
{Mamon}, G.~A., {Glassgold}, A.~E., \& {Huggins}, P.~J. 1988, \apj, 328, 797

\bibitem[{{Morris} \& {Jura}(1983)}]{MJ83}
{Morris}, M. \& {Jura}, M. 1983, \apj, 264, 546

\bibitem[{{Press} {et~al.}(1992){Press}, {Teukolsky}, {Vetterling}, \&
  {Flannery}}]{Press1992}
{Press}, W.~H., {Teukolsky}, S.~A., {Vetterling}, W.~T., \& {Flannery}, B.~P.
  1992, {Numerical recipes in FORTRAN. The art of scientific computing}

\bibitem[{{Ramstedt} {et~al.}(2020){Ramstedt}, {Vlemmings}, {Doan},
  {Danilovich}, {Lindqvist}, {Saberi}, {Olofsson}, {De Beck}, {Groenewegen},
  {H{\"o}fner}, {Kastner}, {Kerschbaum}, {Khouri}, {Maercker}, {Montez},
  {Quintana-Lacaci}, {Sahai}, {Tafoya}, \& {Zijlstra}}]{Ramstedt20}
{Ramstedt}, S., {Vlemmings}, W.~H.~T., {Doan}, L., {et~al.} 2020, \aap, 640,
  A133

\bibitem[{{Saberi} {et~al.}(2019){Saberi}, {Vlemmings}, \& {De
  Beck}}]{Saberi19}
{Saberi}, M., {Vlemmings}, W.~H.~T., \& {De Beck}, E. 2019, \aap, 625, A81

\bibitem[{{Sahai}(1990)}]{Sahai1990}
{Sahai}, R. 1990, \apj, 362, 652

\bibitem[{{Sch{\"o}ier} \& {Olofsson}(2001)}]{Schoier01}
{Sch{\"o}ier}, F.~L. \& {Olofsson}, H. 2001, \aap, 368, 969

\bibitem[{{Visser} {et~al.}(2009){Visser}, {van Dishoeck}, \&
  {Black}}]{Visser09}
{Visser}, R., {van Dishoeck}, E.~F., \& {Black}, J.~H. 2009, \aap, 503, 323

\end{thebibliography}

\end{document}